\documentclass[reprint, amsmath, amssymb, aps, superscriptaddress]{revtex4-2}

\usepackage{graphicx}% Include figure files
\usepackage{dcolumn}% Align table columns on decimal point
\usepackage{bm}% bold math
\usepackage{hyperref}% add hypertext capabilities
\usepackage[caption=false]{subfig}
\usepackage{xcolor}
\usepackage{siunitx}

\newcommand*{\atom}[1]{{\ensuremath{\text{#1}}}} % fuer atomare Symbole
\newcommand*{\molec}[2]{{\ensuremath{\atom{#1}_{#2}}}} % zum Setzen von Molekuelen
\begin{document}

%%%%%%%%%%%%%%%%%%%%%%%%%%%%%%%%%%%%%%%%%%%%%%%%%%%%%%%%%%%%%%%%%
%%%%%%%%%%%%%%%%%%%%%%%%%%%%%%%%%%%%%%%%%%%%%%%%%%%%%%%%%%%%%%%%%
\title{Velocity Map Imaging Spectrometer Optimized for Reduction of Background from Scattered UV Light}

%%%%%%%%%%%%%%%%%%%%%%%%%%%%%%%%%%%%%%%%%%%%%%%%%%%%%%%%%%%%%%%%%

%%%%%%%%%%%%%%%%%%%%%%%%%%%%%%%%%%%%%%%%%%%%%%%%%%%%%%%%%%%%%%%%%

\author{Nicolas Ladda}
\email{Nicolas.Ladda@uni-kassel.de}
 \affiliation{University of Kassel, Institute of Physics, Heinrich-Plett-Str. 40, 34132 Kassel, Germany}
 
\author{Fabian Westmeier}%
 \affiliation{University of Kassel, Institute of Physics, Heinrich-Plett-Str. 40, 34132 Kassel, Germany}

\author{Sagnik Das}%
 \affiliation{University of Kassel, Institute of Physics, Heinrich-Plett-Str. 40, 34132 Kassel, Germany}

\author{Wilfried Dreher}%
 \affiliation{Sch\"ulerforschungszentrum Nordhessen SFN, Parkstrasse 16, 34119 Kassel, Germany}

\author{Simon T. Ranecky}%
 \affiliation{University of Kassel, Institute of Physics, Heinrich-Plett-Str. 40, 34132 Kassel, Germany}
  
\author{Tonio Rosen}%
 \affiliation{University of Kassel, Institute of Physics, Heinrich-Plett-Str. 40, 34132 Kassel, Germany}

\author{Krishna Kant Singh}%
 \affiliation{University of Kassel, Institute of Physics, Heinrich-Plett-Str. 40, 34132 Kassel, Germany}
 
\author{Till Jakob Stehling}%
 \affiliation{University of Kassel, Institute of Physics, Heinrich-Plett-Str. 40, 34132 Kassel, Germany}
 
\author{Sudheendran Vasudevan}%
 \affiliation{University of Kassel, Institute of Physics, Heinrich-Plett-Str. 40, 34132 Kassel, Germany}
  
 \author{Hendrike Braun}%
 \affiliation{University of Kassel, Institute of Physics, Heinrich-Plett-Str. 40, 34132 Kassel, Germany}
 
 \author{Thomas Baumert}%
 \affiliation{University of Kassel, Institute of Physics, Heinrich-Plett-Str. 40, 34132 Kassel, Germany}
 
 \author{Jochen Mikosch}%
 \affiliation{University of Kassel, Institute of Physics, Heinrich-Plett-Str. 40, 34132 Kassel, Germany}
 
  \author{Arne Senftleben}%
\email{arne.senftleben@uni-kassel.de}
 \affiliation{University of Kassel, Institute of Physics, Heinrich-Plett-Str. 40, 34132 Kassel, Germany}
%%%%%%%%%%%%%%%%%%%%%%%%%%%%%%%%%%%%%%%%%%%%%%%%%%%%%%%%%%%%%%%%%

%%%%%%%%%%%%%%%%%%%%%%%%%%%%%%%%%%%%%%%%%%%%%%%%%%%%%%%%%%%%%%%%%
\begin{abstract}
Velocity map imaging spectroscopy is a powerful technique for detecting the momentum distribution of photoelectrons resulting from an ionization experiment on atoms or molecules. However, when used with ultraviolet light sources, scattered photons can lead to emission of photoelectrons from the spectrometer's electrodes, giving rise to severe noise disturbing the desired signal. We present a velocity map imaging spectrometer optimized to reduce such unwanted background signals. The primary modifications to the conventional design include spectrometer electrode geometries with small cross section exposed to the scattered photons, with blocked pathways for photoelectrons from the electrodes to the detector, as well as the incorporation of optical baffles. Compared to a conventional design that is optimized solely on the spectrometer's photoelectron momentum resolution, we have achieved the elimination of 99.9 \% of the background noise without substantial compromise to the resolution. Note that most of the improvements were achieved without the necessity of high-grade windows, reducing the sensitivity to window degradation by the UV light. We give general guidelines on how to efficiently cope with the long-standing experimental problem of electron background originating from scattered light, by considering it already in the design stage of a new spectrometer.

\end{abstract}
%%%%%%%%%%%%%%%%%%%%%%%%%%%%%%%%%%%%%%%%%%%%%%%%%%%%%%%%%%%%%%%%%

%\keywords{Suggested keywords}%Use showkeys class option if keyword
                              %display desired
\maketitle

%\tableofcontents

%%%%%%%%%%%%%%%%%%%%%%%%%%%%%%%%%%%%%%%%%%%%%%%%%%%%%%%%%%%%%%%%%

%%%%%%%%%%%%%%%%%%%%%%%%%%%%%%%%%%%%%%%%%%%%%%%%%%%%%%%%%%%%%%%%%
\section{Introduction}
\noindent
Velocity map imaging (VMI) spectroscopy \cite{Eppink.1997} represents a powerful experimental method for detecting the momentum of charged particles. In this technique, static electric fields are used to map particles with the same charge and momentum vector onto the same point on a detector, independent of their precise initial position in the source volume \cite{Parker.2009}. In the conventional configuration, a position-sensitive detector, typically the combination of a micro-channel plate (MCP), a phosphor screen, and a digital camera, is employed to measure a two-dimensional (2D) projection of the momentum distribution. This scheme works well for distributions which exhibit cylindrical symmetry \cite{Whitaker.2009}. The VMI method is typically applied to photoionization experiments (see below) or collision experiments involving charged particle beams \cite{Mikosch.2006,  Kundu.2023}. Different variants of the conventional scheme are routinely used, particularly for situations without cylindrical symmetry. In slice imaging, the detector is activated only for a short time window, such that exclusively particles with initial momentum parallel to the detector are recorded \cite{Gebhardt.2001, Townsend.2003, Lin.2003}. In tomographic reconstruction, a laser pulse is rotated about its propagation direction and a set of photoelectron angular distributions (PADs) are recorded at different rotation angles \cite{Wollenhaupt.2009b}. In three-dimensional (3D) imaging, time- and position information of particle impact on the detector is determined using two digital cameras with different integration time \cite{Strasser.2000}, cameras with high time-resolution \cite{Zhao.2017, Clark.2012}, the correlation of independently determined camera and time-of-flight information \cite{Wester.2014, Basnayake.2022, Goudreau.2023}, or delay-line or strip anodes instead of a phosphor screen \cite{Vredenborg.2008}. Meanwhile, also double-sided VMI spectrometers are used to record photoelectrons and -ions in coincidence \cite{Rosch.2022}.

The first imaging spectrometer capable of measuring the entire momentum distribution simultaneously was constructed by Chandler and Houston in 1987, utilizing a repeller electrode and a grounded grid electrode \cite{Chandler.1987}. Subsequently, Eppink and Parker incorporated a third electrode (extractor) into the configuration. Their setup was analogous to the time-of-flight spectrometer developed by Wiley and McLaren \cite{Wiley.1955}. However, in contrast to the grid electrodes used in the setup of Wiley and McLaren, the solid electrodes with apertures used by Eppink and Parker introduced inhomogeneous electric fields, significantly enhancing the momentum resolution \cite{Eppink.1997}. After this pioneering work, various researchers have contributed to advancing the technology by adding additional electrodes to improve the resolution \cite{Leon.2013, Leon.2014, Plomp.2021}. Some researchers have optimized the VMI design, enabling the measurement of electrons with high kinetic energy of up to 1 keV \cite{Kling.2014, Li.2018}. Others have implemented Einzel lenses to enlarge the distribution on the detector \cite{Offerhaus.2001, Stodolna.2013}, or optimized the shape of the electrodes to increase the resolution further \cite{Wrede.2001}.

The VMI technique is often used to investigate the momentum distribution of photoelectrons, which can be used to study photoionization processes \cite{Gitzinger.2010, Geiler.2011, Roeterdink.2001}. Combined with femtosecond laser pulses, ultrafast dynamics can be observed \cite{Hansen.2011, Minemoto.2018}. In the context of chiral molecules, VMI is used to measure the Photoelectron Circular Dichroism (PECD), an enantiomer-sensitive effect of the photoelectron momentum distribution \cite{Lux.2012}.

With UV laser pulses, many molecules can be excited from the ground state with one photon and subsequently ionized with a second photon \cite{Liu.2020}. However, photons of sufficiently high energy for such a scheme ($>$ 5 eV) can generate a lot of background noise. In VMI spectroscopy, the light beam usually enters the spectrometer chamber through a window, causing some light to scatter \cite{Horke.2012, Ding.2021}. The scattered photons have enough energy to overcome the work function of metals that are used for the spectrometer electrodes \cite{Ren.2018}, which are often made of aluminium, steel, or copper, all of which have a work function of less than 5 eV \cite{Eastment.1973, Marlow.2023,  Anderson.1949, Skriverand.1992}. Gold plating of the electrodes should result in a surface work function greater than 5 eV. However, the work function of gold is also reduced below 5 eV when exposed to air \cite{Kim.2021}. The work function of other metal-oxides or graphite is also reduced by the exposure to air \cite{Bai.2023, Rietwyk.2019} or surface roughness \cite{Li.2005}.

The usual approach to cope with scattered light is an optical baffle system, an approach inspired by astronomical telescopes \cite{Deilamian.1992}. The optical baffle system aims to block pathways from the window to the surfaces in the spectrometer. However, limitations arise from multiple scattering events at the optical baffle system itself, in particular at grazing incidence, and the sheer number of incident photons. In a single femtosecond laser pulse of \SI{10}{\micro\joule} pulse energy there are around 10$^{13}$ photons of. At the same time the rate of useful events in a femtosecond experiment with gas phase targets is often very low, on the order of one per laser pulse - and these events arrive at the same time on the detector as the background noise. Therefore, scattered light is a long-standing challenging problem in photoelectron spectroscopy involving UV laser pulses.  

Here, we present a velocity map imaging spectrometer (VMIS) that suppresses background electrons resulting from scattered UV light, mostly via a suited geometry of the spectrometer electrodes. We have identified key design considerations that reduce the background, which we present in order of their significance. Note that the electrode material or the vacuum windows are not the main factors in solving the problem of background signal resulting from UV photons.

%%%%%%%%%%%%%%%%%%%%%%%%%%%%%%%%%%%%%%%%%%%%%%%%%%%%%%%%%%%%%%%%%

%%%%%%%%%%%%%%%%%%%%%%%%%%%%%%%%%%%%%%%%%%%%%%%%%%%%%%%%%%%%%%%%%
\section{Optimizations for background reduction}
\noindent
When designing a VMIS, momentum resolution is often the primary concern, and the ability to suppress background from scattered light is neglected. This issue is then addressed later, for example, by using high-quality vacuum windows to minimize scattering \cite{Roberts.2005}. The problem with this approach is that the laser windows often degrade over time, either by the formation of adhesive layers or by damage from the UV laser beam, in particular, color center formation. Our strategies reduce the background noise for a VMIS to acceptable levels, even if the windows are a significant source of scattered photons.

\begin{figure}
    \centering
        \subfloat[]{
        \includegraphics[width=0.45\columnwidth]{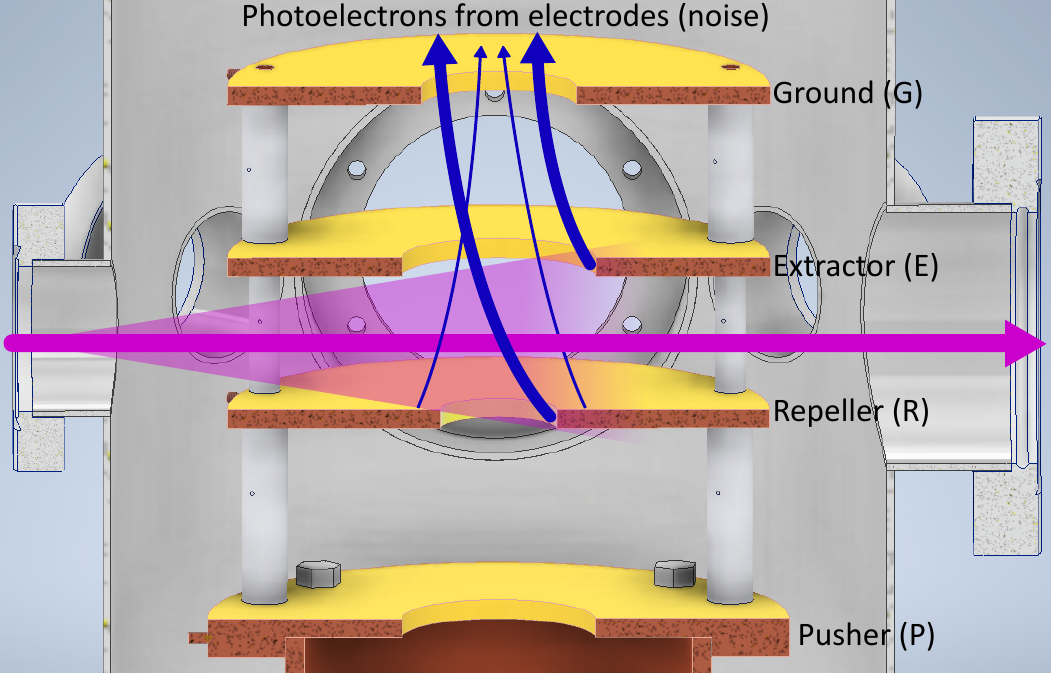}
        \label{VMIS_ThickPlates}
        }
    \hfil
        \subfloat[]{
        \includegraphics[width=0.45\columnwidth]{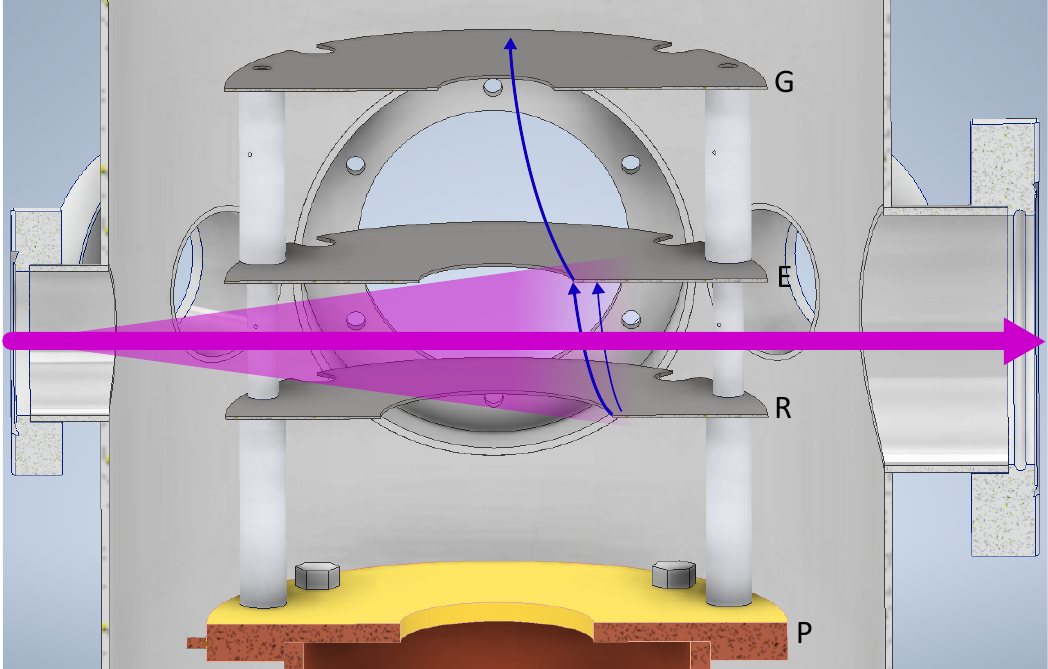}
        \label{VMIS_ThinPlates}
        }
    \caption{Comparison of VMIS with different electrode thicknesses. \protect\subref{VMIS_ThickPlates} shows the resolution-optimized VMIS with thick electrodes, which lead to the creation of a lot of electrons (blue arrows) near the electrode apertures. \protect\subref{VMIS_ThinPlates} shows a VMIS with thin electrodes and a better choice of electrode apertures. Here, due to the smaller cross-section, fewer electrons are created from the electrodes, and most are blocked by the extractor plate, reducing the background signal.}
    \label{VMIS_Design}
\end{figure}

Our VMIS consists of four electrodes: ground, extractor, repeller, and pusher plates, as illustrated in \autoref{VMIS_ThickPlates}. The electrodes are insulated from each other by Teflon rods. The pusher plate was initially installed to enable coincidence detection of ions in a time-of-flight spectrometer but turned out to be crucial for the low background VMIS (see below). The detector comprises a 75 mm diameter, funnel-type MCP \cite{Fehre.2018} with a P43 phosphor screen. The fourth harmonic of a femtosecond titanium-sapphire laser was used as an ionization source. The laser pulses have a central wavelength of 197 nm (6.26 eV photon energy) and a pulse energy of \SI{1}{\micro\joule} at a repetition rate of 3 kHz. The beam diameter and pulse duration are approximately 3 mm and 100 fs, respectively. For this investigation, the unfocussed laser beam was passed through the vacuum chamber without target ($10^{-8}$ mbar vacuum), to assess the background from scattered light. The initial windows were made of low-grade \molec{CaF}{2}, which causes a lot of scattering. For quantitative comparison, images of the background signal were taken under the same conditions, such as MCP voltage, camera exposure time, and number of averaged images. 

\begin{figure}
    \centering
        \subfloat[]{
        \includegraphics[width=0.45\columnwidth]{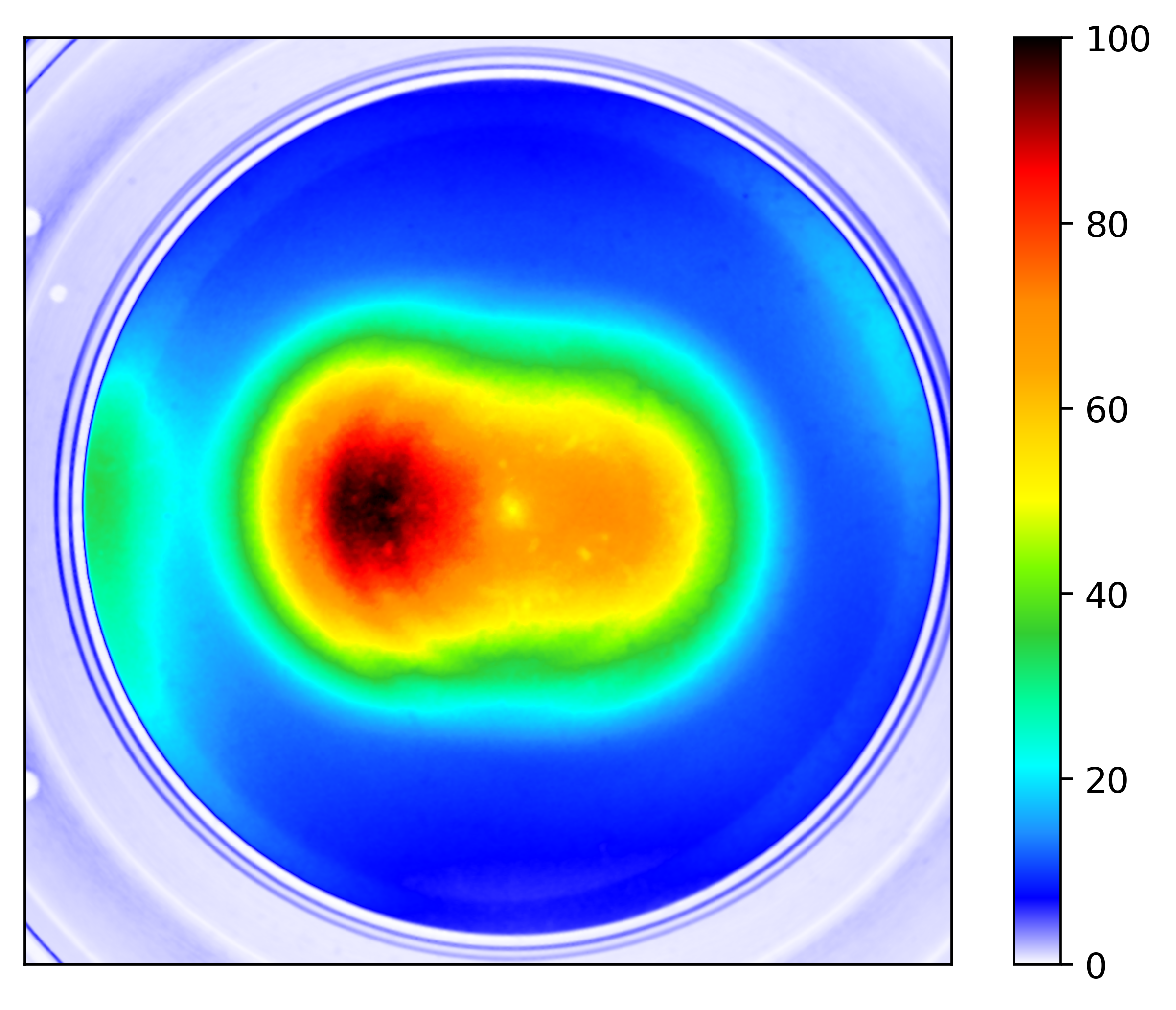}
        \label{VMIS_Design_Pad_Bad}
        }
    \hfil
        \subfloat[]{
        \includegraphics[width=0.45\columnwidth]{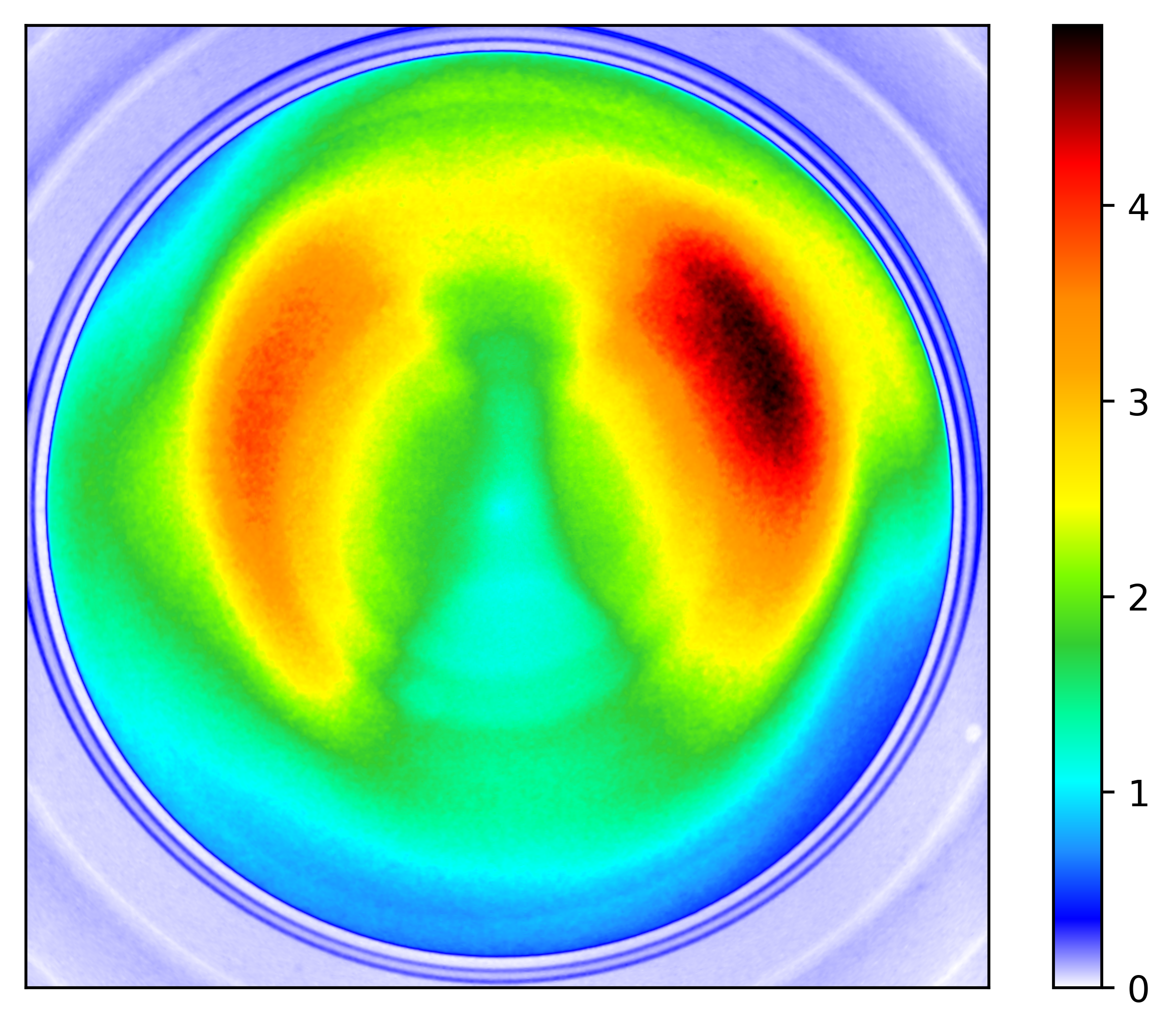}
        \label{VMIS_Design_Pad_Good}
        }
    \caption{Comparison of the background signal measured without target gas under the same conditions between the two VMIS designs in \autoref{VMIS_Design}. \protect\subref{VMIS_Design_Pad_Bad} shows the background signal for the initial resolution-optimized design. \protect\subref{VMIS_Design_Pad_Good} shows the background signal for the improved design, with thinner electrodes and a larger aperture in the repeller than in the extractor.}
    \label{VMIS_Pads_Design}
\end{figure}

We started our investigations with a resolution-optimized VMIS (see \autoref{VMIS_ThickPlates} that was designed with the help of electron trajectory simulations with SIMION 8.0.8 \cite{Dahl.2000}, a standard wide-spread tool for developing ion and electron optics. As in many other VMI spectrometers, our optimization resulted in relatively thick electrodes, in our case 5 mm thick. As a precaution to cope with scattered light, gold-plated copper electrodes were employed since gold should have a higher work function than steel. The resulting background measurement is depicted in \autoref{VMIS_Design_Pad_Bad}. A high, unspecific signal is seen that overshadows any other signal. The visual shape of this signal response to changes in the voltages on the electrodes, therefore we conclude that most of the background consists of electrons. Interestingly, we also observed this background when using the third harmonic of our titanium-sapphire laser with a central wavelength of 263 nm (4.7 eV). The evidence is that a gold surface exposed to air suffers from adsorption, which is not easily removed under vacuum conditions and drastically reduces the work function \cite{Rietwyk.2019, Kim.2021}.

\subsection{Spectrometer design}

We concluded that the inner rims of the thick electrodes constitute the most significant source of background signal in \autoref{VMIS_Design_Pad_Bad}, as the cross-section exposed to the incident light direction is high (see schematic light ray traces in \autoref{VMIS_ThickPlates}). Moreover, light that hits the surface of the electrodes, in particular the top of the repeller electrode, does this under a grazing angle of incidence, which reduces its absorption very significantly. Photoelectrons created at the rims are accelerated straight onto the detector, resulting in severe background noise. Therefore, it is essential that the rim's cross-section is small. This can be achieved by using wedge-shaped electrodes, with the thickness decreasing towards the aperture, or simply thin electrodes. In our improved design, we used plan-parallel thin plates (0.8 mm thick), made of stainless steel (see \autoref{VMIS_ThinPlates}). 
 \begin{figure}
     \centering
         \subfloat[]{
         \includegraphics[width=0.45\columnwidth]{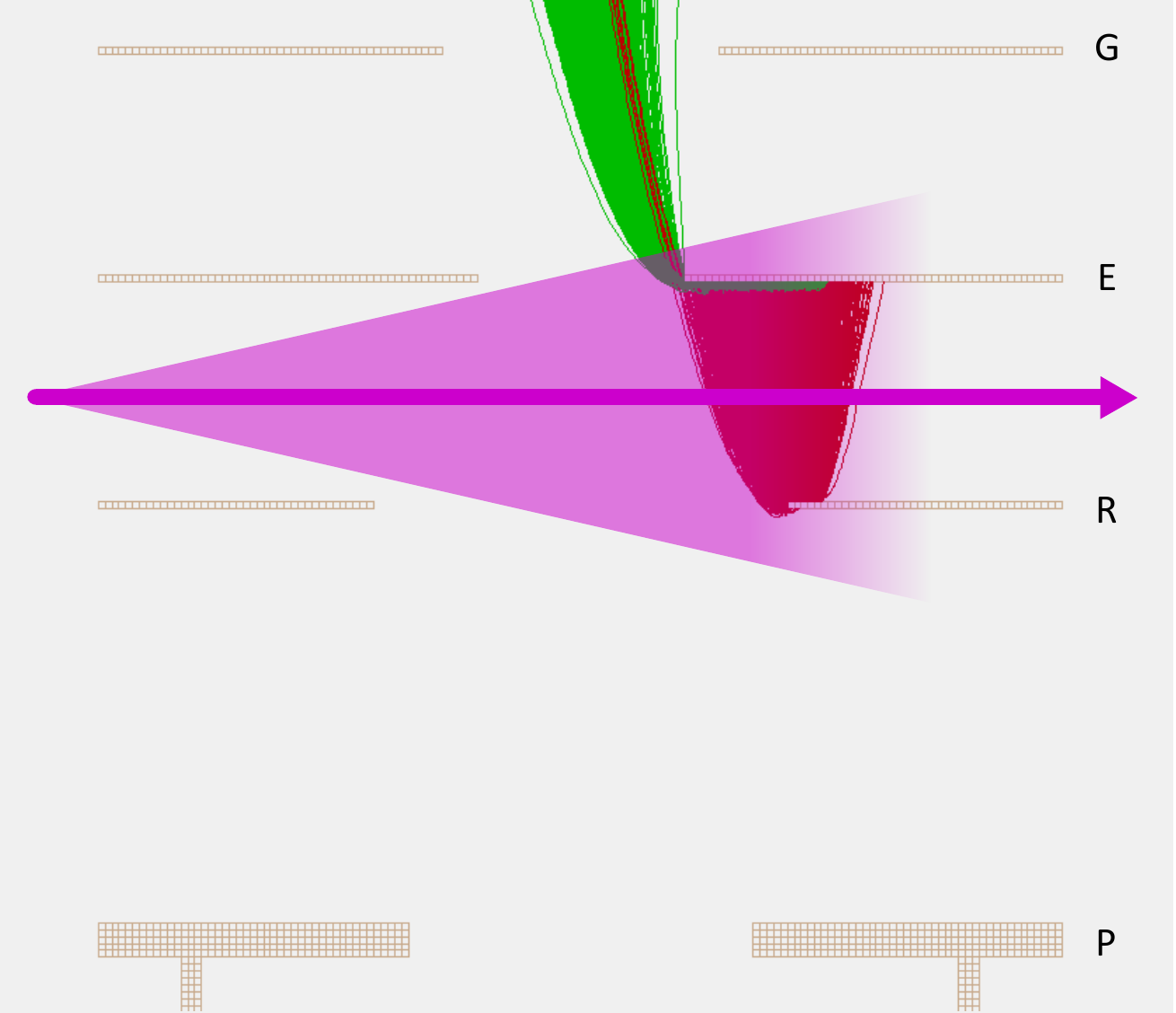}
         \label{Simion_Pusher}
         }
     \hfil
     %\par\bigskip
         \subfloat[]{
         \includegraphics[width=0.45\columnwidth]{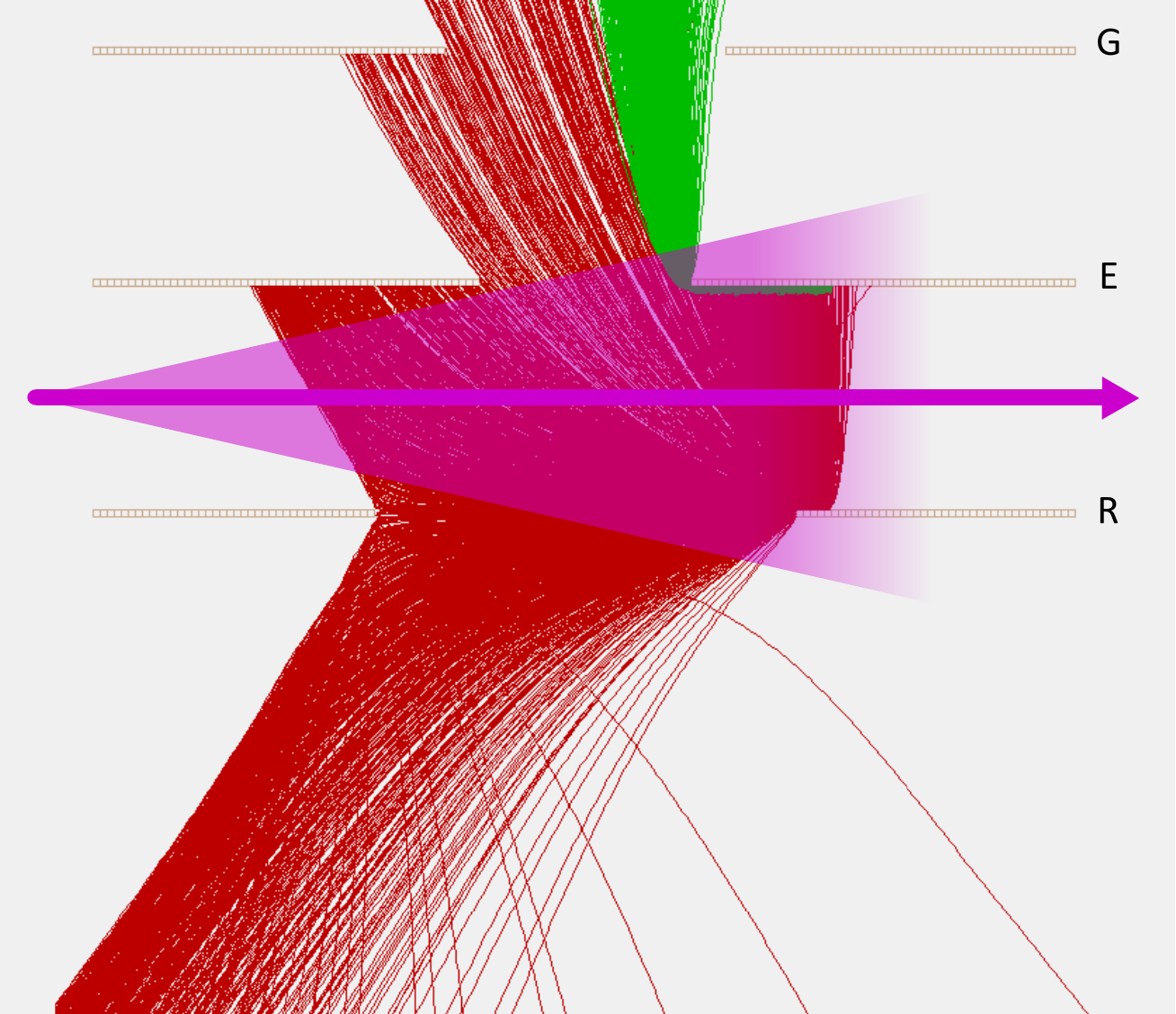}
         \label{Simion_No_Pusher}
         }
     \caption{Comparison of simulated electron trajectories \protect\subref{Simion_Pusher} with and \protect\subref{Simion_No_Pusher} without pusher electrode. The red and green lines represent electrons ejected from the repeller and the extractor plate, respectively. With the pusher electrode (P) more electrons emitted from the repeller plate are directed into the extractor plate, reducing the background signal.}
     \label{Simion_Pusher_Electrode}
 \end{figure}

Furthermore, the electrodes in a VMIS are usually designed such that the aperture diameter in the extractor plate is at least as large as that in the repeller plate. However, to reduce background noise, the opposite is advisable. We use a diameter of 60 mm in the repeller and 40 mm in the extractor plate (see \autoref{VMIS_ThinPlates}). Most electrons emitted from the repeller plate are then blocked by the extractor plate instead of reaching the detector (see \autoref{VMIS_Design}). However, when using a large hole in the repeller electrode, an additional electrode below the repeller plate is crucial. This pusher electrode helps to improve the resolution of the VMIS by a factor of approximately two while also helping to reduce the background signal, according to our SIMION simulations (see \autoref{Simion_Pusher_Electrode}).

The exact distances between the electrodes and the aperture sizes should be optimized for each spectrometer design individually. It is also possible to add additional lenses to increase the momentum resolution. In our improved design of the VMIS, the distance between the pusher and repeller is 65 mm, while the repeller--extractor and extractor--ground distance are 35 mm and 40 mm, respectively. The aperture dimensions of the electrodes are as follows: 50 mm for the pusher, 60 mm for the repeller, 40 mm for the extractor, and 30 mm for the ground plate.

We quantify the improvement by comparing the signal strength of the background noise before and after the implementation of each design modification. The following data processing procedure is conducted in three steps: First, the minimum value of each image is determined and subsequently subtracted from each pixel. Then, the value of each pixel is divided by the maximum value across all images. Finally, the result is multiplied by 100. Consequently, the image of the initial design had pixel values ranging from 0 to 100 (\autoref{VMIS_Design_Pad_Bad}). Images of the adapted designs had values between 0 and $<100$ (\autoref{VMIS_Design_Pad_Good}), demonstrating the extent to which background noise is reduced with each optimization. Notably, after implementing the improvements in the electrode design, the highest pixel value was less than 5, indicating that more than 95\% of the background noise has successfully been removed (see \autoref{VMIS_Pads_Design}).

\subsection{Optical baffles}

\begin{figure}
    \centering
        \subfloat[]{
        \includegraphics[width=0.3\columnwidth]{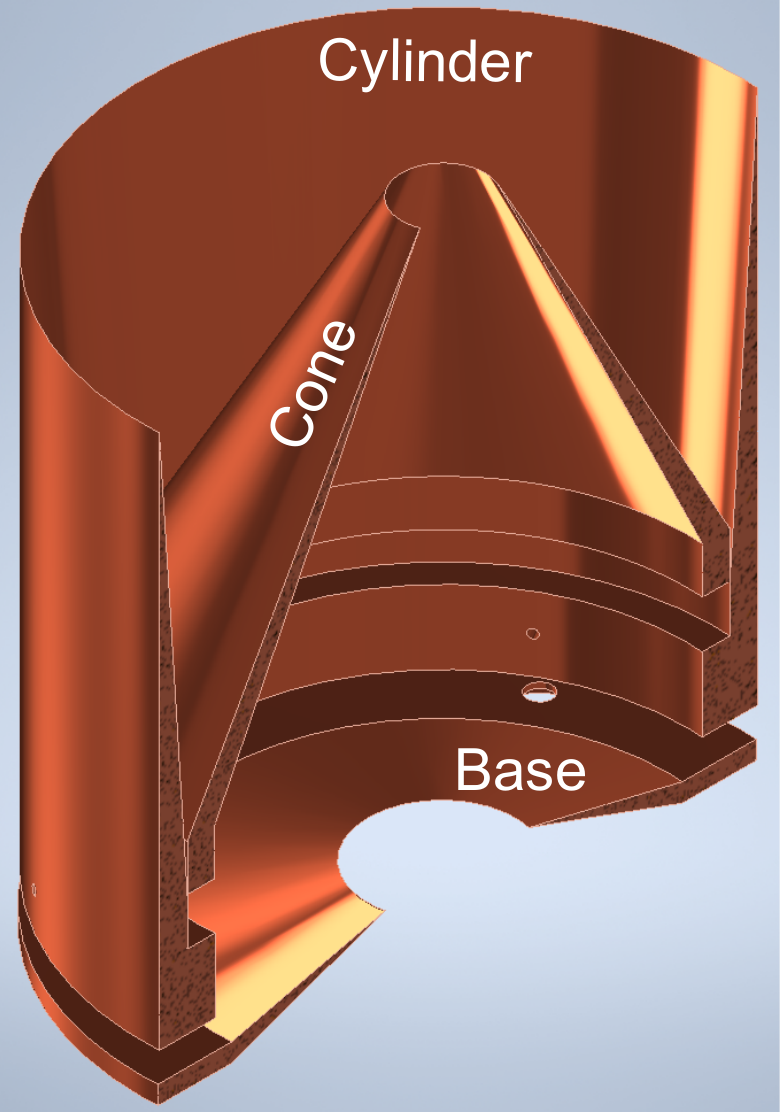}
        \label{Baffle_Copper_Design}
        }
    %\hfil
        \subfloat[]{
            \includegraphics[width=0.3\columnwidth]{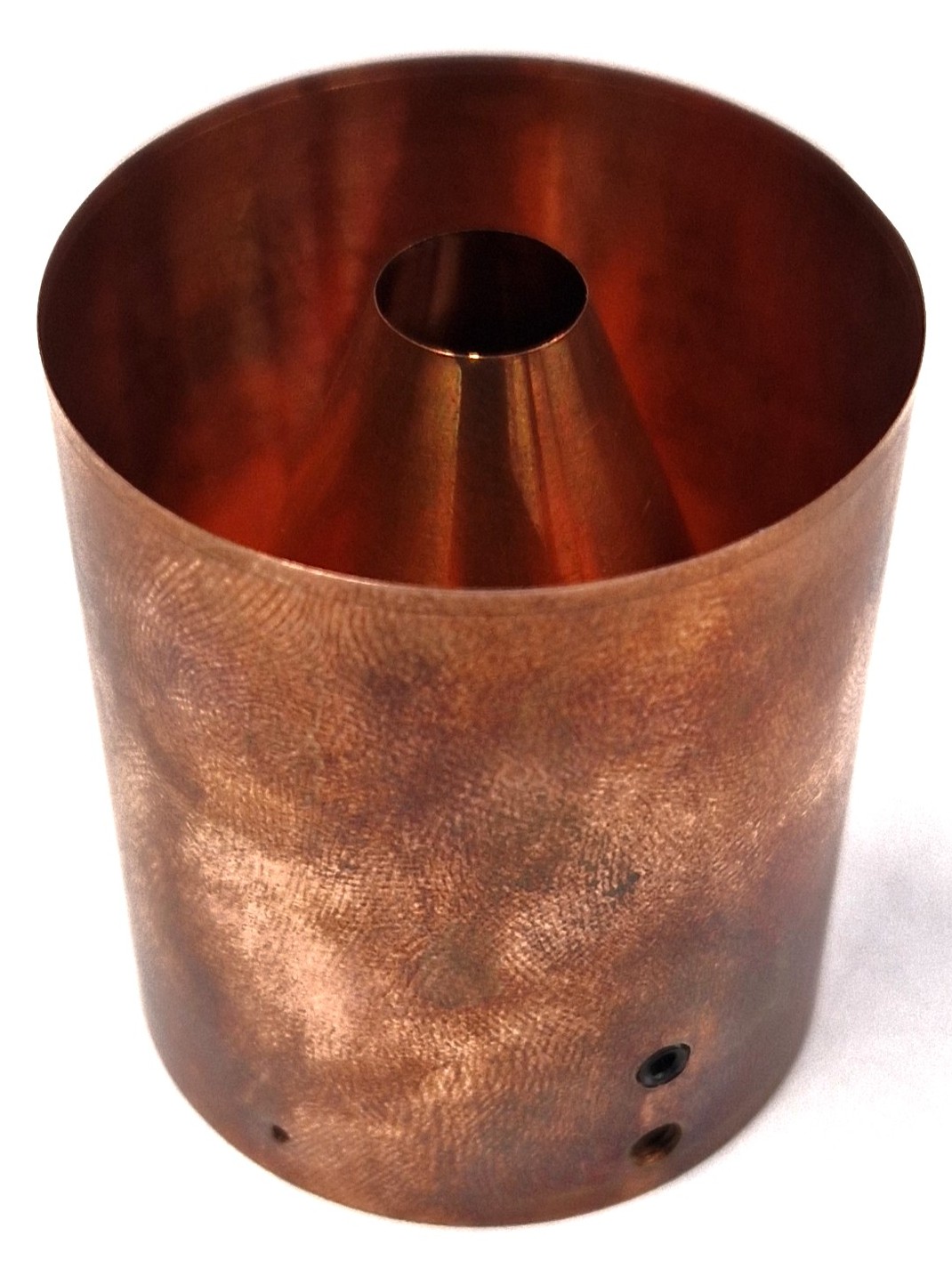}
            \label{Baffle_Copper}
            }
    %\hfil
        \subfloat[]{
        \includegraphics[width=0.3\columnwidth]{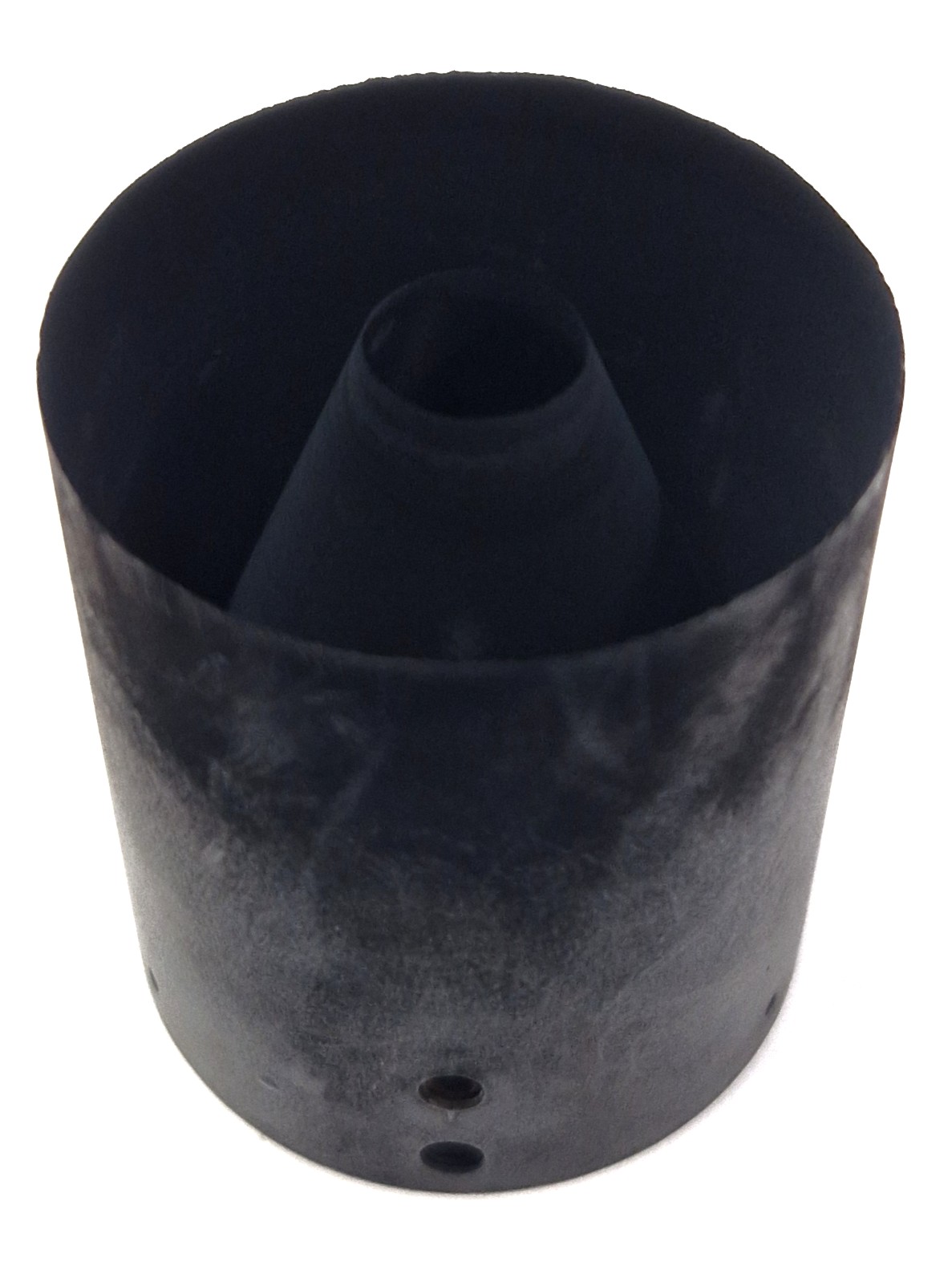}
        \label{Baffle_Black}
        }
    \caption{Modular design of the optical baffles \protect\subref{Baffle_Copper_Design}, consisting of a cylinder, a conical piece and a base plate. optical baffles before \protect\subref{Baffle_Copper} and after the blackening \protect\subref{Baffle_Black} process.}
    \label{Baffles}
\end{figure}

The purpose of optical baffles is to largely block the path of scattered light between its source, in our case, the vacuum window transmitted by the laser, and the sensitive surface, the spectrometer electrodes. Importantly, since multiple scattering can occur, it is important to find a good compromise between blocking scattered light and not cutting too much into the wings of the laser beam, which typically has a Gaussian spatial profile. We use a modular, vacuum-compatible design for the optical baffle system. One element consists of three pieces: a cylinder, a cone, and a base plate (see \autoref{Baffle_Copper_Design}, where the assembly is shown). The edges of the cone are as thin as possible (knife-edge) to minimize specular reflection from the edge towards the spectrometer.

We usually operate our VMIS in a back-focussed arrangement, in which the collimated laser beam is transmitted through the vacuum chamber via an entrance and exit window and then back-reflected by a concave mirror placed in front of the exit window, which produces a focus in the source volume of the VMIS. This arrangement has advantages over other configurations: (1) Due to the focusing from a reflecting surface, there is no chromatic effect. Two collimated laser beams with different wavelengths are focussed at the same position, which is important for multi-color pump-probe experiments. (2) No astigmatism is introduced, and a shorter focal length is possible compared to folded beam paths. (3) By placing the focussing mirror outside the vacuum, it is easily accessible for alignment and replacement.    

\begin{figure}
    \centering
        \subfloat[]{
        \includegraphics[width=0.45\columnwidth]{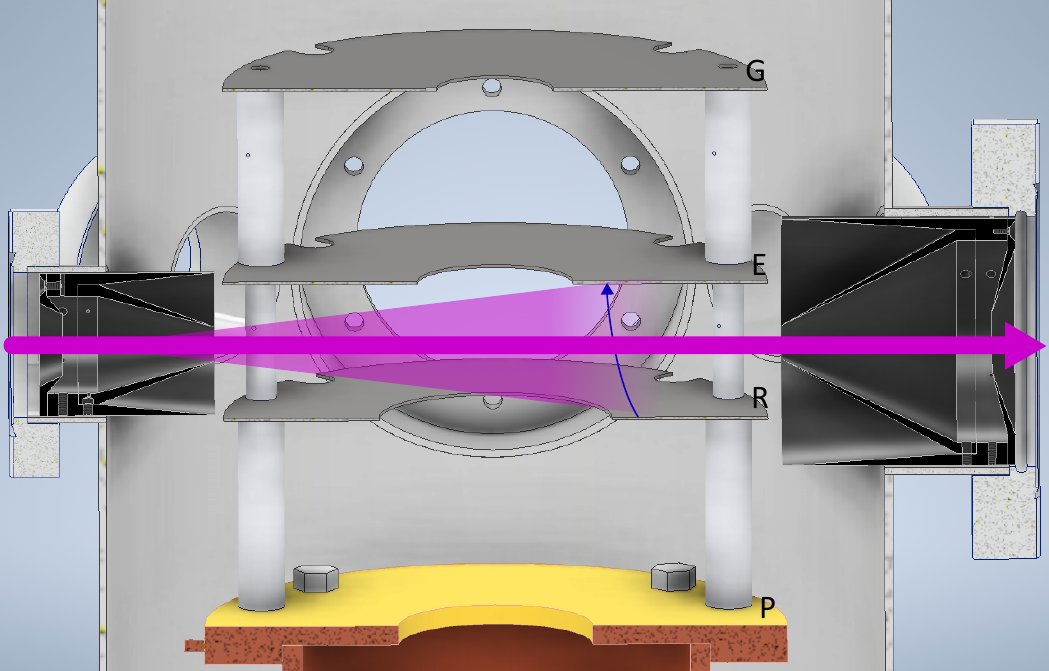}
        \label{VMIS_Design_Baffles}
        }
    \hfil
        \subfloat[]{
        \includegraphics[width=0.45\columnwidth]{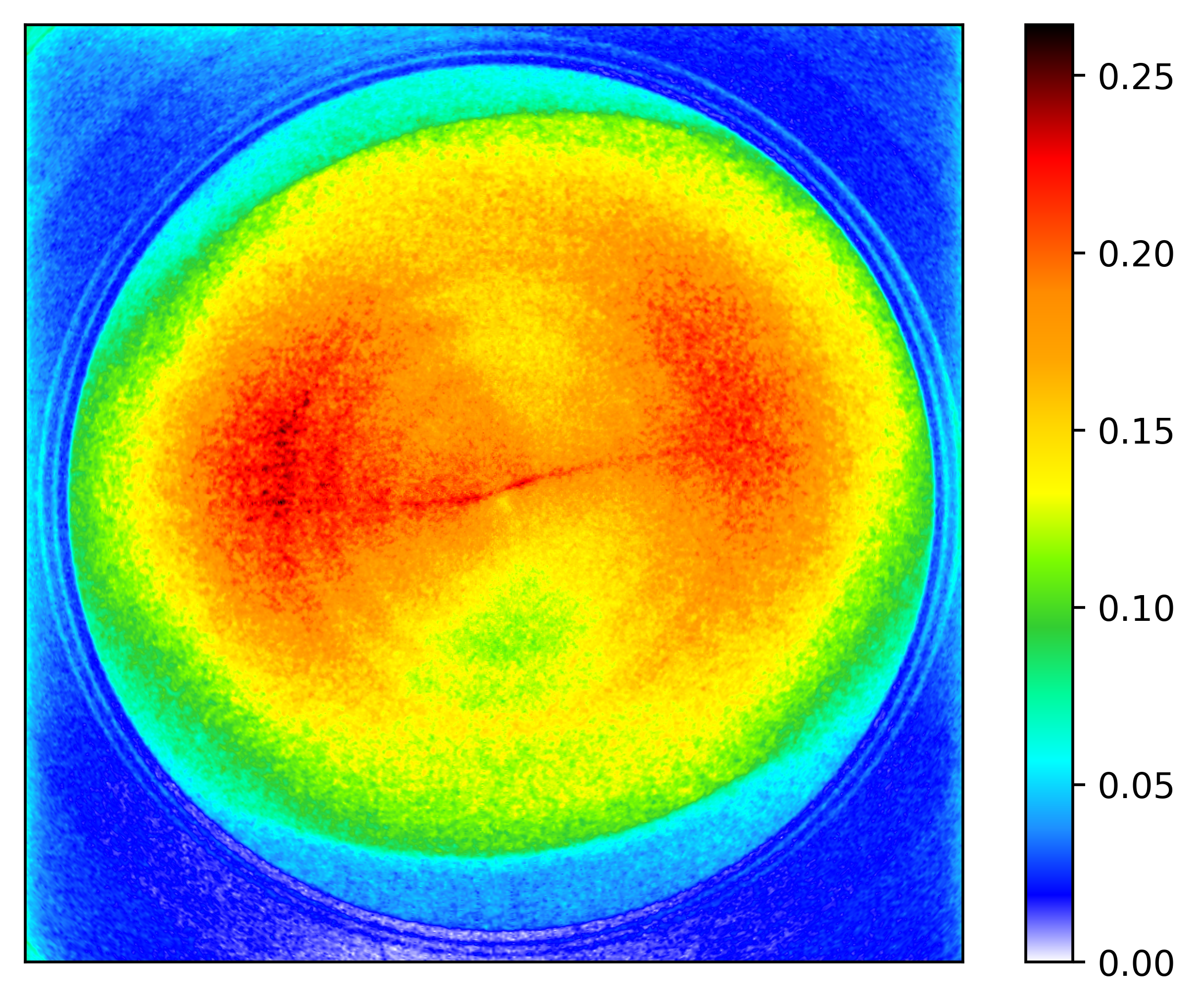}
        \label{VMIS_PAD_Baffles}
        }
    \caption{Influence of the optical baffles on the background noise. \protect\subref{VMIS_Design_Baffles} shows the spectrometer with the optical baffles installed, which confine the solid angle of scattered light entering the spectrometer. \protect\subref{VMIS_PAD_Baffles} shows the background signal for this design.}
    \label{VMIS_Design_Bad_Vs_Good}
\end{figure}

To block scattered light both from the incoming and the back-focussed beam, one of the optical baffle elements described above (see \autoref{Baffle_Copper_Design}) is placed on each side of the VMIS stack, as seen in \autoref{VMIS_Design_Baffles}. The cones are facing inwards to bring their aperture as close as possible to the interaction regime and confine the solid angle of scattered light as much as possible.

The optical baffles must exhibit high absorption in the UV and be conductive enough to avoid charge build-up, to not disturb the spectrometer fields. A material that fulfills these criteria is copper covered by a thin layer of cupric oxide (CuO) with a dendritic surface structure \cite{OwenJ.Clarkin.2012}. A wet etching technique can easily create the oxide layer. The following procedure was adapted from \citeauthor{OwenJ.Clarkin.2012} \cite{OwenJ.Clarkin.2012}.

First, it is necessary to remove any larger dirt particles from the copper parts with water and soap. For the primary cleaning step, acetone and isopropanol should be used. Afterward, the pieces should be soaked in concentrated acetic acid at approximately 50°C for 3--5 minutes. This will remove most residual oxide layers from the surface without affecting the copper. It is advised that these two steps are repeated until the surface appears thoroughly cleaned, ending with the acetic acid bath.

In the next step of the process, the copper pieces are etched with nitric acid (20\%) at a temperature of 50°C for a duration of between 60 and 90 seconds. This step removes any residual oxides and roughens the surface of the pieces. Subsequently, the copper pieces are washed several times in distilled water to remove all traces of nitric acid. It is essential to note the importance of this step, which must be carried out with the utmost care, as any residual acid can have a significantly detrimental effect on the quality of the final result. It is essential to employ a two-water bath system, with the second water bath being slightly basic (with the addition of NaOH). It is crucial to progress directly to the final step after etching to avoid the re-formation of oxide layers, which would also compromise the quality of the results.

In the final step of the process, the etched copper pieces are oxidized by potassium persulfate (20 g/L) in a 60 g/L NaOH solution at 55°C. It is critical to maintain a temperature below 60°C and to ensure that stirring is not excessive. The reaction is self-limited and typically takes 25 minutes to complete. After additional washing and drying in a drying cabinet, the outcome of this reaction is a surface that is characterized by its extremely high degree of blackness and its high UV absorption capabilities \cite{Richharia.1990, OwenJ.Clarkin.2012}.

\begin{figure}
	\centering
	\includegraphics[width=0.9\columnwidth]{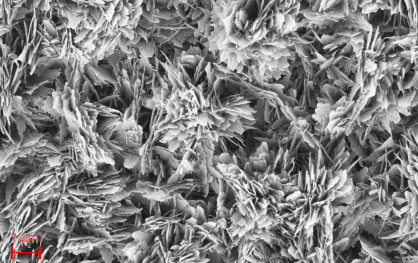}
	\caption{Surface structure of the blackened optical baffle imaged with an SEM. The surface was coated with a minimal amount of gold to increase secondary electron yield. The dendritic cupric oxide crystals are less than 100 nm thin and oriented randomly.}
	\label{Baffle_SEM}
\end{figure}

We examined the cupric oxide surface by scanning electron microscopy (SEM). \autoref{Baffle_SEM} presents SEM images of the surface structure of the blackened optical baffles. To enhance the secondary electron yield, a minimal amount of gold was sputtered onto the surface. A comparison with unsputtered samples revealed that this process did not affect the underlying structure. It is evident that the surface is composed of thin (less than 100 nm) but elongated cupric oxide crystals, which are oriented in an irregular pattern. This dendritic structure facilitates light absorption with high efficiency \cite{Richharia.1990, OwenJ.Clarkin.2012}. 

In \autoref{VMIS_Design_Bad_Vs_Good}, the VMIS design incorporating optical baffles and the corresponding background noise measurement are presented. The confinement of the solid angle of scattered light leads to a reduction in the electrode surface area exposed to scattered photons, consequently diminishing the number of photoelectrons near the electrode apertures. This requires a sufficient distance between the repeller and the extractor electrode. In the case of the VMIS presented here, this gap is 35 mm wide, while the electrodes themself have a diameter of 140 mm. Additional SIMION simulations showed that the grounded optical baffles can harm the resolution of the VMIS if the electrode plates are too small in diameter, reducing their ability to block the altered electrostatic field by the optical baffles. 

Our measurements under defined conditions (see above) show that the incorporation of optical baffles in combination with the described geometrical modifications of the spectrometer electrodes result in a substantial reduction of background signal, with about 99.75\% of the noise being suppressed as compared to the initial design (see \autoref{VMIS_PAD_Baffles}). The laser beam must be aligned precisely through the optical baffle system to minimize scattering originating from clipping.

\subsection{Windows}

The utilization of high-quality vacuum windows is beneficial in reducing the amount of scattering, with calcium fluoride ($\mathrm{CaF_2}$) and magnesium fluoride ($\mathrm{MgF_2}$) being particularly effective materials due to their high UV transmission. For $\mathrm{MgF_2}$, its birefringent nature poses a challenge in experiments that demand precise control over light polarization. We took care that the crystal purity is high to minimize both scattering and the degradation of the windows. In addition, it is recommended that the surfaces of the crystal be polished to a low roughness in order to reduce scattering further.

We use 1.3 mm thick, 25 mm diameter optically polished VUV-grade single crystal $\mathrm{CaF_2}$ as vacuum windows (Korth Kristalle), which we embedded with UHV compatible glue into a recess in a CF40 flange. We took care to align the laser beam as parallel as possible to the surface normal of the windows. This was done to minimize detrimental back reflections at an angle \cite{Roberts.2005}.

\begin{figure}
    \centering
    \includegraphics[width=0.45\columnwidth]{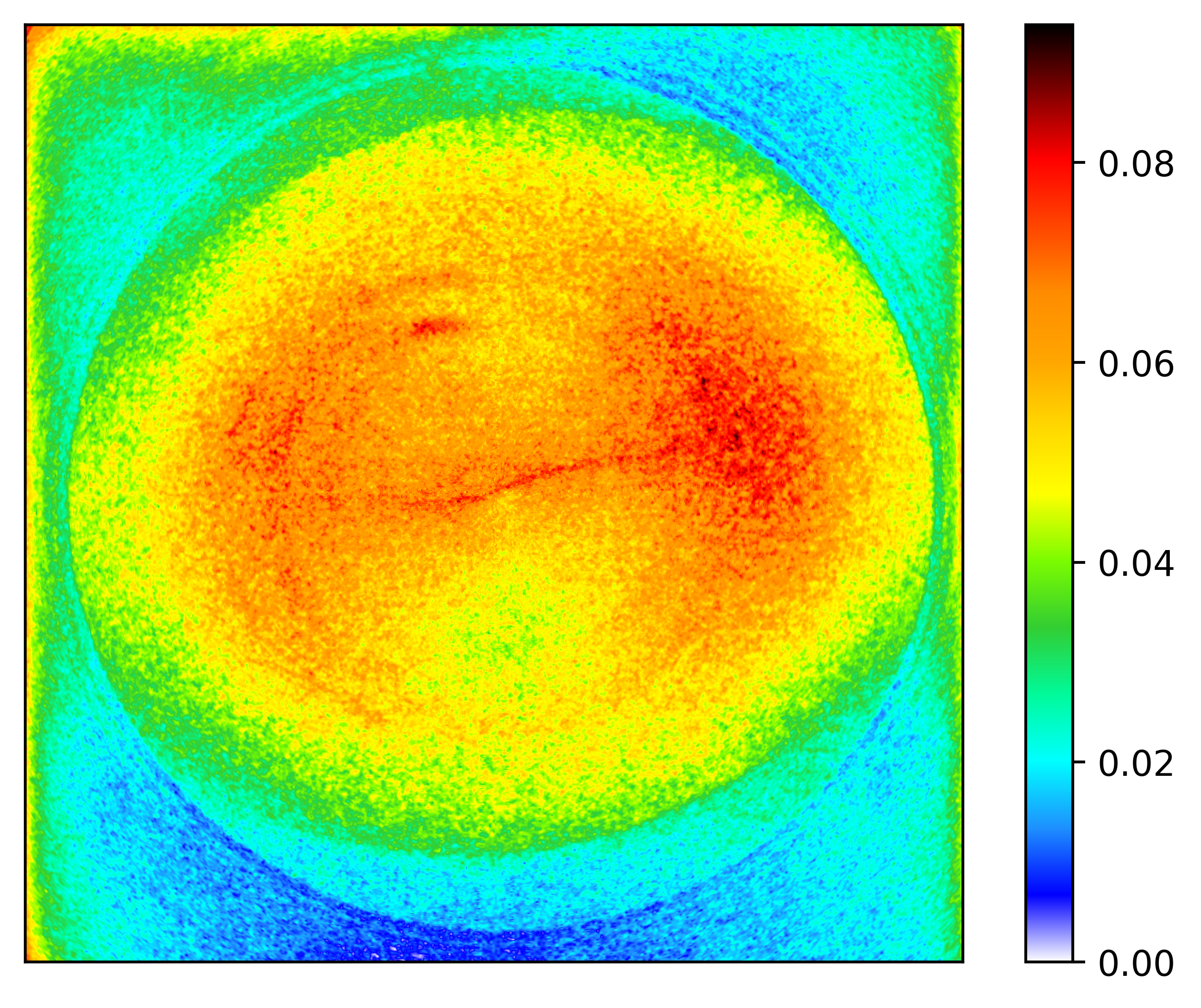}
    \caption{Background signal after all improvements (electrode design, optical baffles, and high-grade vacuum windows). The Background level has been reduced by more than a factor of 1000 compared to the initial design (compare \autoref{VMIS_Design_Pad_Bad})}
    \label{VMIS_Baffles}
\end{figure}

With the high-quality windows added to the modifications described above, we eliminate 99.9 \% of the initial background signal (see \autoref{VMIS_Baffles}). Please note that the improvement is only a factor of three compared to the poor-quality windows, indicating that the windows are no longer the dominant source of background electrons from scattered UV photons with the geometrical design adaptations described above. This is especially interesting since windows are prone to degradation over time due to laser-induced damage or accumulation of contamination on the surfaces.

%%%%%%%%%%%%%%%%%%%%%%%%%%%%%%%%%%%%%%%%%%%%%%%%%%%%%%%%%%%%%%%%%

%%%%%%%%%%%%%%%%%%%%%%%%%%%%%%%%%%%%%%%%%%%%%%%%%%%%%%%%%%%%%%%%%
\section{Results}
\noindent
\begin{figure}
    \centering
        \subfloat[]{
        \includegraphics[width=0.45\columnwidth]{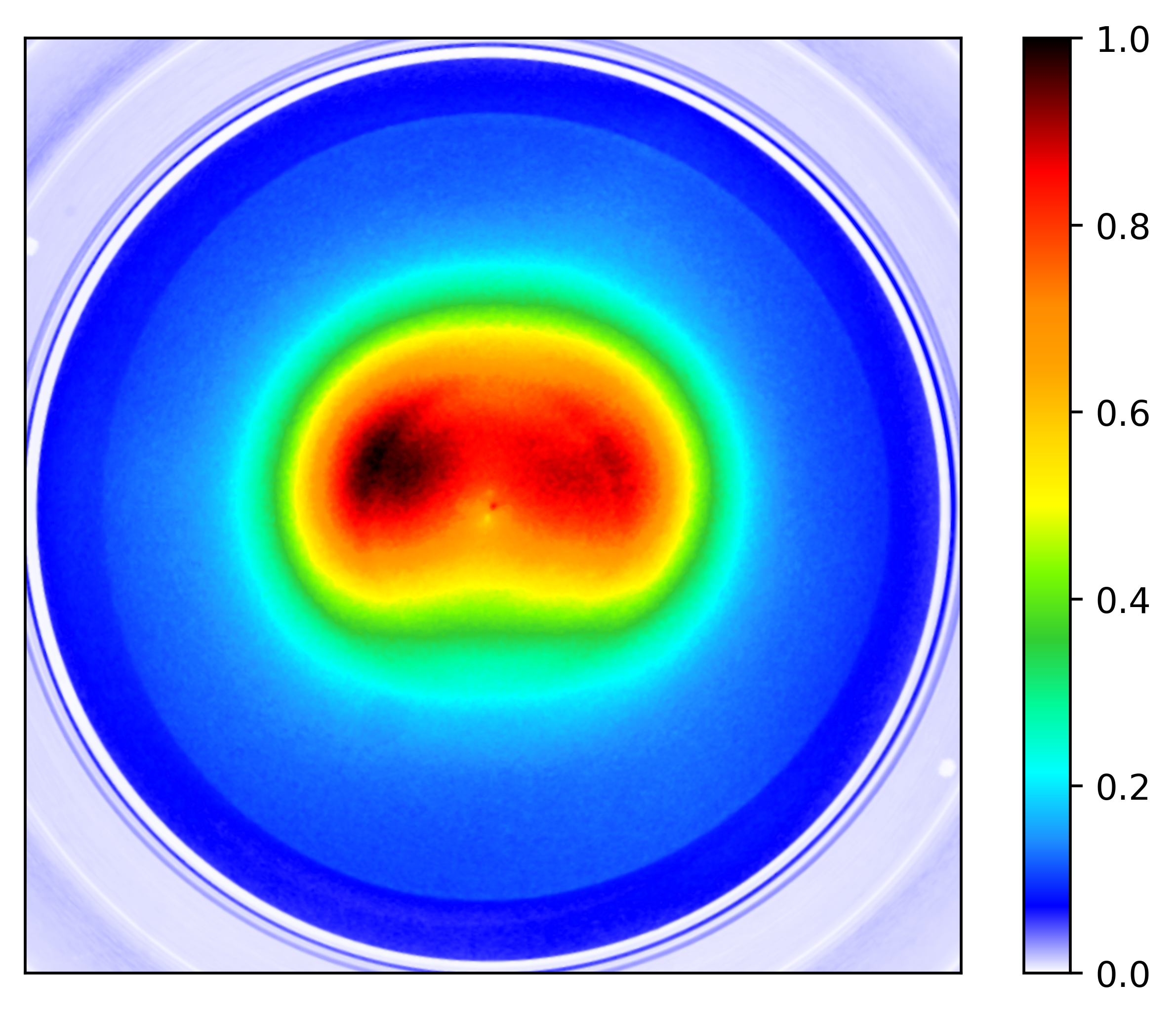}
        \label{PAD_Fenchone_Old_VMIS}
        } 
    \hfil
        \subfloat[]{
        \includegraphics[width=0.45\columnwidth]{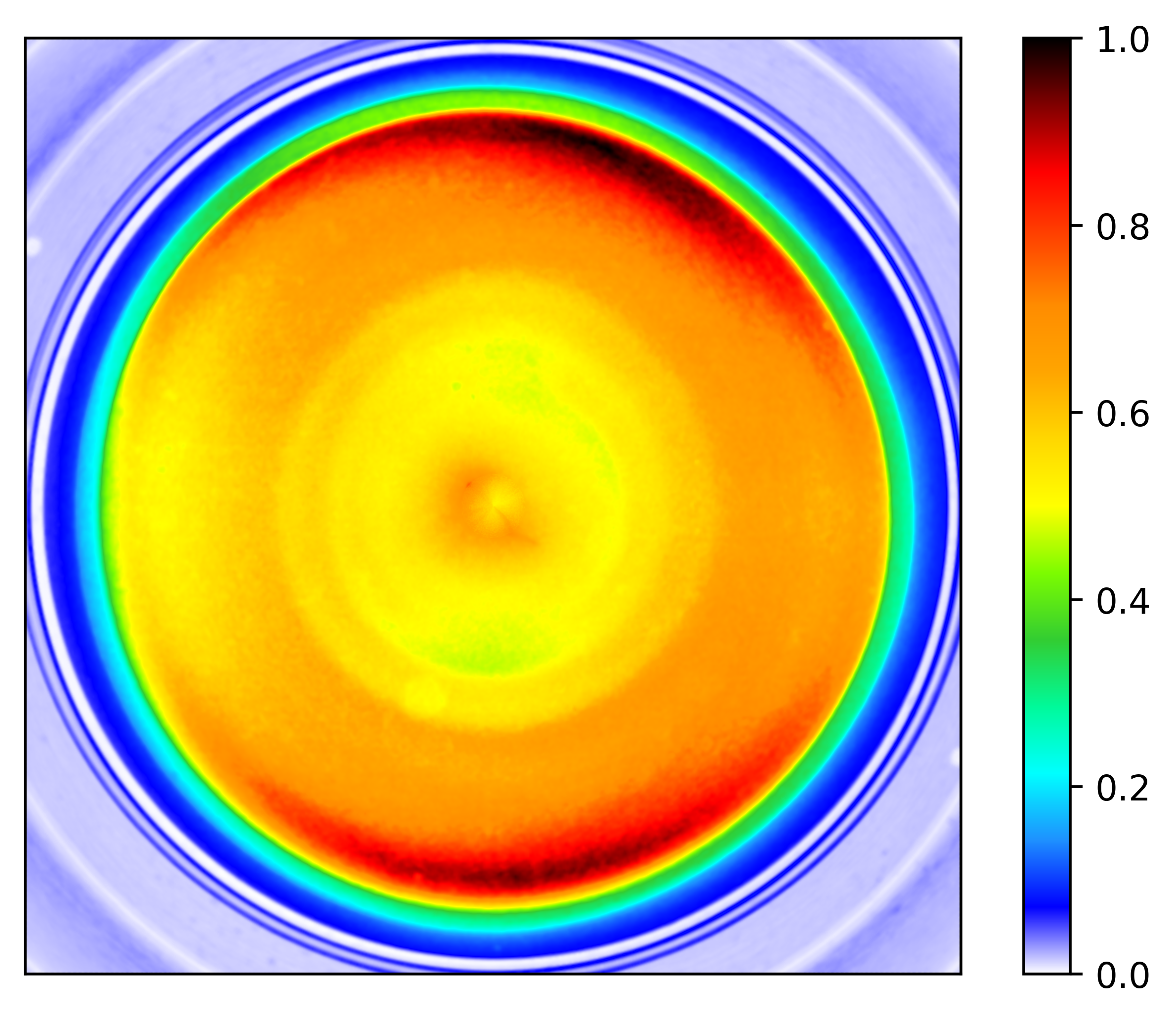}
        \label{PAD_Fenchone_New_VMIS}
        }
    \caption{Comparison between measurements of fenchone taken with the fourth harmonic (198 nm) before \protect\subref{PAD_Fenchone_Old_VMIS} and after \protect\subref{PAD_Fenchone_New_VMIS} implementing the changes to reduce background signal.}
    \label{PAD_Fenchone}
\end{figure}

The improved VMIS featured a background signal that was more than 1000 times lower than that of the initial design. \autoref{PAD_Fenchone} presents a comparison of a 1+1 Resonance-Enhanced Multi-Photon Ionization (REMPI) photoelectron momentum distribution of fenchone with 197 nm wavelength femtosecond pulses. The left image illustrates the recorded data obtained from the previous spectrometer design, in which a pronounced background signal is prominently visible in the centre. The primary signal is barely perceptible. In contrast, the image on the right was recorded using the novel setup, where virtually no background signal is visible.

% ich finde das mit dem fenchone muss unbedingt bleiben. wenn es nicht genau der alte aufbau war, kann man das mit einzelnen wörtern abfedern

\begin{figure}
    \centering
        \subfloat[]{
        \includegraphics[width=0.45\columnwidth]{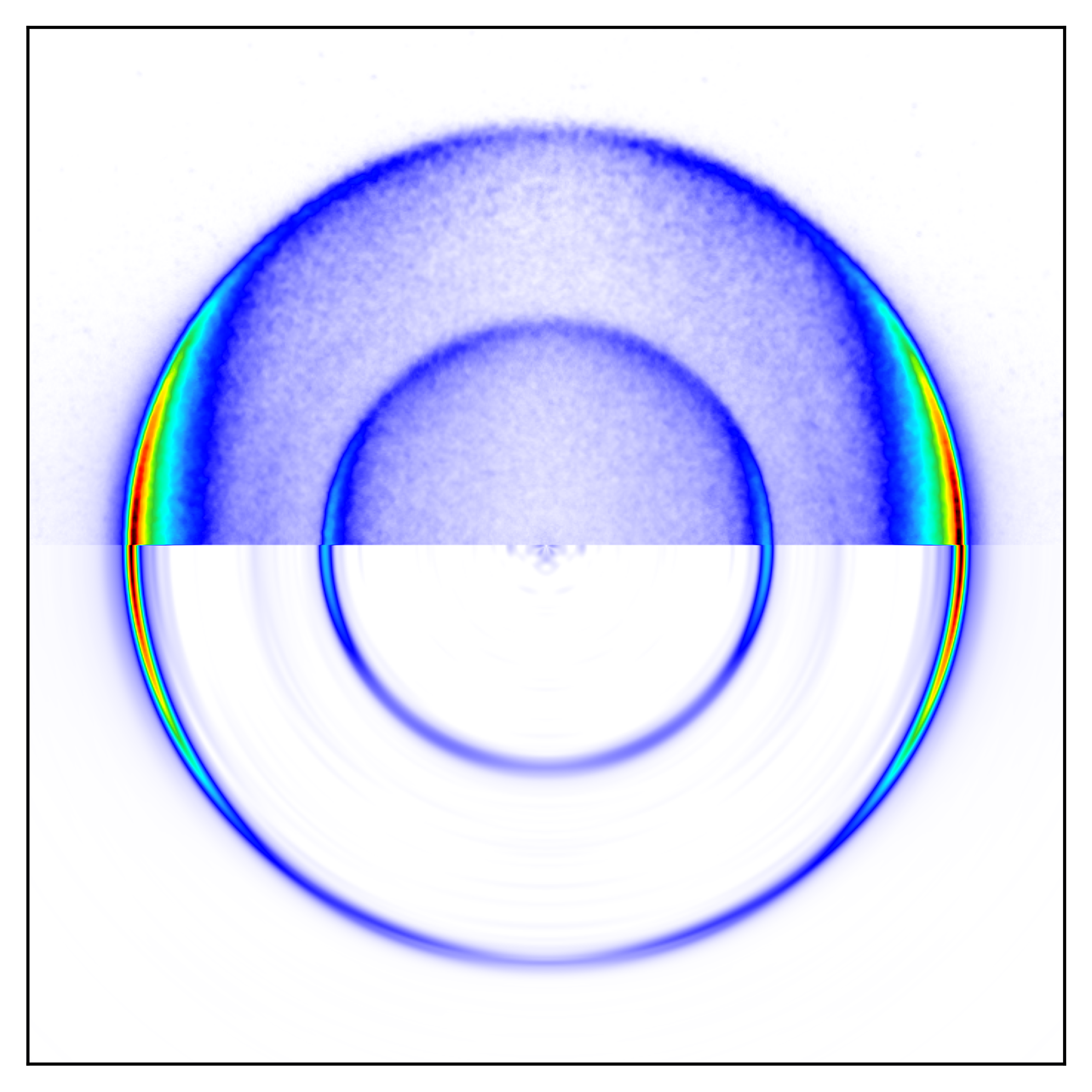}
        \label{PAD_Xenon}
        }
    \hfil
        \subfloat[]{
        \includegraphics[width=0.45\columnwidth]{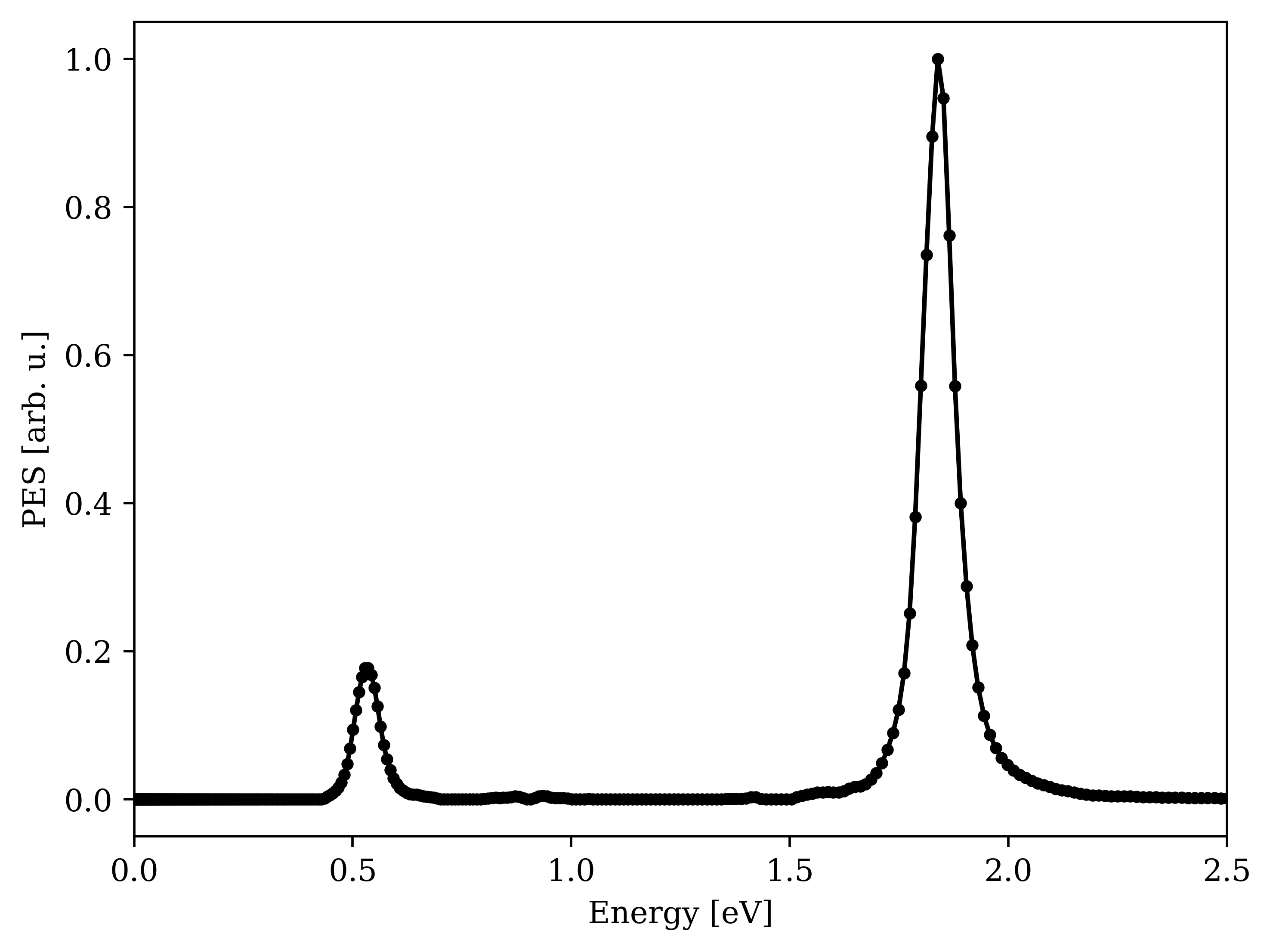}
        \label{PES_Xenon}
        }
    \caption{Photoelectron momentum distribution of xenon \protect\subref{PAD_Xenon} and the extracted photoelectron kinetic energy spectrum \protect\subref{PES_Xenon}, as recorded with the third harmonic of a nanosecond YAG laser (355 nm) with the background-optimized VMIS. Decent momentum resolution was found. The upper half of the image in \protect\subref{PAD_Xenon} shows the raw data, while the lower half represents the Abel-inverted data.}
    \label{Xenon}
\end{figure}

To check the resolution of the new VMIS, photoelectrons from multi-photon ionization of xenon with the third harmonic of a nanosecond Nd:YAG laser (355 nm wavelength) were measured. The recorded detector image (2D projection of the photoelectron momentum distribution) and the corresponding photoelectron spectrum are depicted in \autoref{Xenon}. The measured resolution for the background-optimized VMIS design is $\Delta E_{FWHM}/E = 4.7 \%$ at a kinetic energy of 1.9 eV, comparable to the previous resolution-optimized design. The FWHM of the Abel-inverted signal (not shown) is approximately 7 pixels. A single event generates a signal with an FWHM of approximately 4 pixels, indicating that the VMIS is operating close to the resolution limit of the detector. Consequently, the design adaptations made to minimize background signal do not significantly compromise the resolution of the VMIS.

% was machte diese Referenz hinter 7 pixels ? \cite{Garcia.2004, MikhailRyazanov.2012}

\bigskip
%%%%%%%%%%%%%%%%%%%%%%%%%%%%%%%%%%%%%%%%%%%%%%%%%%%%%%%%%%%%%%%%%

%%%%%%%%%%%%%%%%%%%%%%%%%%%%%%%%%%%%%%%%%%%%%%%%%%%%%%%%%%%%%%%%%
\section{Conclusion}
\noindent
We present a VMIS design optimized to reduce photoelectron background originating from scattered UV photons. We found the largest potential for background reduction in the geometry optimization of the spectrometer plate electrodes and the installation of an optical baffle system. Specifically the utilization of thin electrodes and the blocking of background electrons by the extractor are recommended. Installation of high-grade laser windows at the vacuum interface can reduce the background signal further, but its effect becomes comparatively small when the other measures are implemented. We have experimentally demonstrated that the background-optimized VMI design features only a small reduction in momentum resolution as compared to a design that is solely optimized on the latter. The background signal of the final design was found to be more than 1000 times lower than that of the resolution-optimised VMIS.

%Eine sache noch: Seht ihr auch UV Photonen direkt? Das sollte man noch hinschreiben. Gibt es die und sie stören nicht, oder gibt es die gar nicht. Was macht uns sicher, daß der Hintergrund Elektronen sind und nicht Photonen? 
%%%%%%%%%%%%%%%%%%%%%%%%%%%%%%%%%%%%%%%%%%%%%%%%%%%%%%%%%%%%%%%%%

\begin{acknowledgments}
Financial support from Deutsche Fors\-chungs\-gemeinschaft (DFG, German Research Foundation) – Projektnummer 328\-96\-1117 – SFB 1319 ELCH is gratefully acknowledged. JM gratefully acknowledges support from the DFG within a Heisenberg professorship (project number 470414645).
\end{acknowledgments}

%apssamp
\bibliography{Literatur/VMIS_Lit}% Produces the bibliography via BibTeX.

\end{document}